\documentclass[12pt]{article}

\usepackage{amsfonts,amsmath,amssymb}

\author{Andrei Khrennikov\\International Center for Mathematical Modeling \\
in Physics, Engineering, Economics, and Cognitive Science\\
Linnaeus University, V\"axj\"o-Kalmar, Sweden }

\title{Measurement problem: from De Broglie to theory of classical random fields interacting
with threshold detectors}

\begin{document}

\maketitle

\begin{abstract}
The quantum measurement problem as was formulated by von
Neumann in 1933 can be solved by going beyond the operational quantum
formalism. In our ``prequantum model" quantum systems are
symbolic representations of classical random fields. The Schr\"odinger's
dynamics is a special form of the linear dynamics of classical fields.
Measurements are described as interactions of classical fields with detectors.
Discontinuity, the ``collapse of the wave function", has the
trivial origin: usage of threshold type detectors. The von Neumann
projection postulate can be interpreted as the formal mathematical
encoding of the absence of coincidence detection in measurement on a
single quantum system, e.g., photon's polarization measurement. Our
model, prequantum classical statistical field theory (PCSFT), in combination
with measurements by threshold detectors satisfies the quantum
restriction on coincidence detections: the second order coherence
is less than one (opposite to all known semiclassical and classical feld
models). The basic rule of quantum probability, the Born's rule, is
derived from properties of prequantum random felds interacting with
threshold type detectors. Comparison with De Broglie's views to
quantum mechanics as theory of physical waves with singularities is
presented.
\end{abstract}

\section{Introduction}

The measurement problem in quantum mechanics [1] is the unresolved problem
of how (or if) wave function collapse occurs. The inability to observe
this process directly has given rise to different interpretations of quantum
mechanics and poses a key set of questions that each interpretation must
answer. (See [2]-[7] for hot debates on the  ``right interpretation".) The
wave function in quantum mechanics evolves according to the Schr\"odinger
equation which preserves linear superposition of different states, but actual
measurements always find the physical system in a definite state. Any future
evolution is based on the state the system was discovered to be in when the
measurement was made, meaning that the measurement ``did something" to
the process under examination. Whatever that ``something" may be does
not appear to be explained by the basic conventional quantum theory.

During the last 12 years quantum foundations were discussed at the series
of conferences which took place in V\"axj\"o (South-East Sweden), see, e.g., [2]-[7]. And the most exciting spectacle started each time when the question
of interpretations of the wave function attracted the attention. Finally, it
became clear that the number of different interpretations is in the best case
equal to the number of participants. If you meet two people who say that
they are advocates of, e.g., the Copenhagen interpretation of QM, ask them
about the details. You will see immediately that their views on what is
the Copenhagen interpretation can differ very much. The same is true for
other interpretations. If two scientists tell that they are followers of Albert
Einstein's ensemble interpretation, ask them about the details... At one of
the round tables (after two hours of debates with opinions for and against
completeness of QM) we had decided to vote on this problem. Incompleteness
advocates have won, but only because a few advocates of completeness voted
for incompleteness. The situation is really disappointing: the basic notion of
QM has not yet been properly interpreted (after 100 years of exciting, but
not very productive debates).

We can also mention the V\"axj\"o interpretation of QM. This interpretation
evolved essentially from the V\"axj\"o interpretation-2000 [8] which was based
on ``naive Einsteinian realism" (the values of physical observables can be
assigned to a quantum system before measurement) to the V\"axj\"o interpretation
2009 in which measurement context played a fundamental role [9] (in
particular, the values of physical observables cannot be assigned to a quantum
system before measurement). The latter combines the views of Einstein
on incompleteness of QM with the views of Bohr on contextual structure of
quantum measurements. By the V\"axj\"o interpretation subquantum reality
exists. However, for a moment we cannot approach it by means of available 
observables. The presently used observables, ``quantum observables",
are not ``elements of subquantum reality": for example, the components of
the electric and magnetic fields of photon or the density of electron's charge,
see W. Hofer [10] for the latter. In our model the presently used ``quantum
observables" are contextual: their values cannot be assigned to subquantum
systems in advance, before measurement.\footnote{This is a good place to point to a general 
scientific methodology which was advertised
during many years by Atmanspacher and Primas [11]. Any scientific theory is based on
two levels of description of reality: ontic (reality as it is) and epistemic (the image of
reality obtained with the aid of a special class of observables). The QM-formalism is an
example of an epistemic model. However, existence of an epistemic model does not prevent
scientists to go beyond it to approach the ontic level.}
 The results of quantum measurements
are determined by measurement contexts. This is really surprising,
because typically contextuality was considered as a quantum feature. The
basic classical physical model, classical statistical mechanics, is not contextual,
the values of physical observables are considered as the values of an
object. However, as the reader will see in the present paper, already classical
wave models combined with measurement theory based on usage of threshold
type detectors can exhibit contextual features.

This is well known (starting with the work of S. Gudder [12], see also K.
Svozil [13] and the author's book [9]) that contextual models can violate Bell's
inequality [14]. Therefore this is not surprising that PCSFT in combination
with measurements of the threshold type can peacefully coexist with Bell's
no-go theorem. (This problem will not be considered in the present paper;
see, however, [15] for the detailed presentation.)\footnote{See [16], [17], [18] for recently constructed prequantum models which do not contradict
to the Bell's argument. We do not comment these models in the present paper.}

There is a plenty of approaches to solve the measurement problem by
using the formalism of quantum mechanics, e.g., [19], see also [20] for the
detailed review. However, we speculate that it seems to be impossible to
find the ``real solution of this problem" in the standard quantum framework,
since the quantum formalism (as was pointed by N. Bohr on many occasions,
e.g., [21], [22]) is a formal operational formalism describing measurements
for microsystems and not physical processes in the microworld [22]: ``Strictly
speaking, the mathematical formalism of quantum mechanics and electrodynamics
merely offers rules of calculation for the deduction of expectations
pertaining to observations obtained under well-defined experimental conditions
specified by classical physical concepts", cf. [23], [24], [25]. By using
the operator representation of observables we escape a detailed description
of the process of measurement and such a description could not recovered
in the quantum formalism. To solve the measurement problem of quantum
theory, one has to go beyond the operational quantum formalism, cf. [19],
[20], [26].

In this paper, we ``solve the quantum measurement problem''\footnote{Of course, one staying at the Copenhagen position would not consider our 
quantum solution of the measurement problem as the real solution of this problem.} by interpreting
quantum systems as formal operational representations of classical
prequantum fields (waves) and by describing the measurement process as
the process of interaction of a classical wave with a threshold type detector.
In section 2 we analyze the process of measurement of a classical wave by
a detector which selects one of eigenfunctions from the coherent superposition
representing the wave, see (3). After such a measurement this superposition
is ``collapsed". This consideration is motivated by chapter 5 of L. De Broglie's
book [27], pp. 51-54; see also [28]. The ``collapse" of a sound wave (emitted
e.g., by a vibrating string) which interacts with the turning fork having
one of the basic string's frequencies can be operationally described in the
same way as the collapse of the wave function of a photon (emitted in the
state of superposition of a few frequencies) which interacts with a photo-detector. 
However, opposite to the quantum wave function, the classical
wave is collapsed not to a single eigenfunction of the corresponding stationary
equation, but to the decoherent mixture of such eigenfunctions. Thus the
formal mathematical description is similar to the description of quantum
measurement without selection of the fixed value of the observable. In the
later case we also obtain not a single eigenvector, but a mixture, described
by the density operator.

Roughly speaking in our approach the main difference between classical
and quantum collapses is degeneration of the former, the impossibility to
select a single output component of the signal (``quantum particle"). In quantum
theory this problem is known as the problem of coincidence detection. It
was intensively studied experimentally to reject (semi)classical models pretending
to reproduce quantum predictions, see Grangier et al. [42], [41],
also [43]. The quantum prediction that coincidence detections should occur
relatively rarely was supported by experiments. It is commonly accepted
that these experiments demonstrated that classical or semiclassical models
without coincidence detection do not exist.

However, the field model elaborated in [29]-[36], prequantum classical statistical
field theory (PCSFT), in combination with usage of the threshold type
and properly calibrated detectors solves the coincidence detection problem,
opposite to known (semi)-classical models. In [37] we found that quantum
statistics can be obtained in a simple way: by combination of stationary
random field describing spatial or internal degrees of freedom with the Brownian 
motion (Wiener process) describing temporal fluctuations, cf. with
[15], where a more complicated stochastic processes were in use.
In this paper we essentially improved the rigorousness of the presentation of
the threshold detection scheme for random waves by coupling this problem
with the well studied problem of probability theory, namely, the first hitting
time problem. Unfortunately, probabilistic studies on hitting times were restricted
to real valued stochastic processes. (These studies were essentially
stimulated by applications to finances and here the real valued processes
are in usage.) Development of theory of hitting times for complex valued
stochastic processes is a complicated mathematical problem and we hope
that our paper will stimulate research in this direction. For a moment, we
can apply only results on hitting times for real processes and therefore we
restrict our consideration to the case of real Hilbert space. In this paper we
reproduce quantum statistics for density matrices with real elements. This is
merely a question of the mathematical justification; in [37] we proceeded in
the general case of complex Hilbert space, but not in the completely rigorous
mathematical framework.

We remark that the aforementioned classical-quantum analogy between
measurements of classical and quantum waves is valid only for disturbative
classical measurements. It is commonly postulated that in classical physics
it is possible to perform measurements with an arbitrary precision; in particular,
it is possible to determine the form of the wave without to destroy
coherent superposition. In this paper, we do not criticize this postulate\footnote{One of the main objections is the presence of noise; for prequantum classical fields
their coupling with the background field, vacuum fluctuations, is irreducible [29]--[36]. For
a moment, we ignore this problem.}. We
just remark that quantum observables form only a subclass of observables
for prequantum fields. These are coarse and disturbabtive measurements.
PCSFT's class of observables is essentially larger and it contains measurements
of fields components, e.g., the electric and magnetic fields components
of a photon. The modern experimental technology is still far from realization of
such measurements, cf., however, with recent experimental results presented
by W. Hofer (Conference ``Quantum Mechanics as Emergent Phenomenon",
Vienna, November 2011) on a possibility to violate Heisenberg's uncertainty
relation.

At the very end of section 2, we compare the De Broglie's Double Solution
theory [28], [27] with PCSFT (combined with threshold-type measurements)
and with Bohmian mechanics. There is a rather common opinion that the
De Broglie's Double Solution theory is simply an early version of a more
advanced theory, namely, Bohmian mechanics. However, the careful study of
works of De Broglie shows that this viewpoint to the inter-relation between
the {\it De Broglie's Double Solution theory} and {\it Bohmian mechanics} is rather
primitive, see section 2.

\section{Classical fields: superposition, linear dynamics, measurement}

In this section we consider classical waves with linear dynamics (in vacuum
or some media). The process of measurement induces effects\footnote{The interaction between a 
measurement device and the input wave can be linear; so
nonlinearity is not crucial for collapse.} which formally
can be described by the von Neumann projection postulate: linear
superposition ``collapses". We claim that the essence of the ``collapse" of
superposition is the transition from one linear dynamics {in the absence of
measurement { to another dynamics (nonlinear or even linear) of interaction
with a measurement device.

A possibility to form linear superposition of waves (fields) and to split a
wave into superposing summands is the basic feature of the classical wave
theory. A crucial point is the existence of dynamics which preserve linear
superpositions in the process of evolution. We can mention the dynamics of
the string or the classical electromagnetic field (Maxwell's equations)\footnote{We remark that all these linear dynamics are approximate. Although these are very
good approximations, one should not overestimate the role of linear dynamics. At the
fundamental level the majority of processes are nonlinear, cf. De Broglie [28].
}.

Suppose that in the absence of interaction with measurement devices the
field dynamics is described by the differential equation:
\begin{equation}
\label{E1}
\gamma\frac{d \phi(t, x)}
{dt}
= L\phi(t, x), 
\end{equation}
where $L$
is a linear partial differential operator and $\gamma$ is a constant. Since, at
each instant of time $t,$ the field's energy is finite,
$\int_{{\bf R}^3} \vert \phi(t, x)\vert^2 dx < \infty,$ the equation
(1) can be considered as a linear (ordinary) differential equation in the $L_2$-
space. This space has the Hilbert space structure. Denote it by the symbol
$H.$ For all basic physical processes, we can assume that the operator $L$
is
self-adjoint in $H.$

Sometimes it is convenient to proceed with complex vector valued fields,
e.g., for the electromagnetic field, we can consider the Riemann-Silberstein
representation, $\phi(t, x) = E(t, x) + iB(t,x),$ where  $E$ and $B$ are the electric and magnetic
components, respectively. We remark that the system of the Maxwell
equations can be written in such a form with  $\gamma= i$ [30], i.e., in the same
form as the Schr\"odinger equation [38], [30]. (We also mention the work of
Strocchi [39] who demonstrated that the Schr\"odinger equation can be written
as the system of linear Hamiltonian equations, see also [30].) In general, we
work with real valued vector fields. The wave equation and Klein-Gordon
equation cannot be written in the complex form, here fields are real and
 $\gamma= 1.$ Solution of a linear dynamical equation with self-adjoint operator $L$
can be represented in the form of superposition of solutions $\phi_k = \phi_k(x)$ of
the stationary equation:
\begin{equation}
\label{E2}
L\phi_k(t, x)= \omega_k \phi_k(t, x); 
\end{equation}
here
\begin{equation}
\label{E3}
\phi(t,x) = \sum_k c_k(t) \phi_k(t, x), c_k(t)= e^{\frac{\omega_k t}{\gamma}} c_{k0},
\end{equation}
where
\begin{equation}
\label{E4}
\phi(0, x) = \sum_k c_{k0} \phi_k(x)
\end{equation}
is the expansion of the initial wave. The main statement is that this form of
expansion, (4), is preserved in the process of evolution. If, for some $k,$ the
term with $\phi_k(x)$ was present in (4) at $t = 0,$ then it will never disappear from
(3), for any instant of time $t > 0.$ (If the operator $L$
has continuous spectrum,
then the sign of sum is changed to the sign of integral.)

As was pointed out by L. De Broglie [27], the expansion (3) is a formal
mathematical representation. The field $\phi(t, x)$ cannot be imagined as
``physical superposition" of fields $\phi_k(x).$ He stressed that the contributions
of summands $\phi_k$ in (3) can be extracted from the field $\phi$ only through interaction
with measurement devices, cf. Roychoudhuri [40].

In [27] the example of vibrating string was presented. The waves $\phi_k$
correspond to the basic frequencies $\omega_k$ of the vibrating string (for some type
of boundary conditions). However, until we start measurement the waves $\phi_k$
do not present in the integral wave $\phi$ so to say  physically. In this case measurement
can be done with the aid of a turning fork. Put a turning fork nearby the
vibrating string and by adjustment of turning fork\footnote{We can use 
either a turning fork which frequency can be changed or what is may be
even easier a collection of turning forks representing scale of frequencies.} we can find one of the
basic frequencies, say $\omega_{k_0}$ . The sound emitted by turning fork at the frequency
$\omega_{k_0}$ corresponds to extracting from the integral wave $\phi$ its fixed component
$\phi_{k_0}.$  Formally, this process can be described as the orthogonal projection $P_{k_0}$
(in Hilbert space $H)$ of the integral wave $\phi$ onto the one dimensional subspace
corresponding to the wave $\phi_{k_0}.$  One can call this process the collapse of the
classical wave $\phi$ or more precisely the collapse of the linear superposition
(3). Of course, there is nothing mysterious in this collapse; in particular,
it is completely clear that it is not instantaneous, the process of interaction
with the turning fork has a finite duration; neither mystery is destruction
of superposition (3): the dynamics of interaction with the turning fork is
different from the dynamics (1) preceding interaction and hence the original
superposition need not be preserved and it can be destroyed, collapsed. (The
interaction dynamics need not be nonlinear. It can be linear as the original
dynamics (1), but with another linear operator, say $L_1,$
Dynamics with $L$
and $L_1$ can be unified through linear dynamics with time dependent generator.)

\medskip

In fact, the situation is more complicated. Here the string plays the role
of a source of sound waves (cf. with a source of quantum systems). Hence,
the turning fork interacts not directly with the string, but with the emitted
sound wave. To proceed rigorously, we have to use a different notation for this
(sound) wave, say $\varphi(t, x)$ and the corresponding solutions of the stationary
equation $\varphi_k(x).$ Before measurement, the dynamics of the sound wave $\varphi(t, x)$
can be well described by the linear partial differential equation, the wave
equation. Hence, as well as (3), the representation
\begin{equation}
\label{E5}
\varphi(t,x) = \sum_k c_k(t) \varphi_k(x), 
\end{equation}
can be used for mathematical calculations. Here we stress again that the
physical (sound) wave $\varphi(t,x)$ is not composed of 
physical stationary (sound) waves $\varphi_k(x).$

\medskip

To simplify consideration, we shall proceed with original vibrations of the
string (by having in mind that, in fact, the measurement device turning fork
destroys the coherent superposition (5) in the sound wave). We state again
that one has to be very careful with terminology. The expression ``destroys
coherent superposition" is related to the ``mathematical wave" and not the
physical one. (Hence, ``collapse" takes place in mathematical space.)
In the presented considerations we stressed the analogy between ``collapses"
of the classical wave and the quantum wave function. The similarity
of operational descriptions of ``collapses", i.e., by using projection operators
in Hilbert space, is especially important. However, although classical superposition
(3) collapsed as the result of interaction with the turning fork which
is used for measurement and in the operational formalism the extraction of
the component $\phi_{k_0}(x)$ can be described by the projection operator b $P_{k_0},$  the
reader experienced in quantum theory would point out that there is a crucial
difference between the collapse of the quantum wave function and the classical
wave. In the classical case, although the component $\phi_{k_0}(x)$ is extracted
from the signal, the signal is not completely reduced to this component. If
we put two turning forks, we can select two basic frequencies, $\omega_j,  
j = 1, 2.$ The selection of each component $\phi_j$ can be operationally represented by the
corresponding projector $P_{j}$ in $H.$ However, opposite to the quantum case
(we consider the case of nondegenerate spectrum, $\omega_i \not= \omega_j, i\not= j)$ the output
signal is not reduced to one of the selected components. And if we put many
(and even infinitely many) turning forks, it is possible to select corresponding
components of the wave $\phi.$

By the conventional interpretation of quantum mechanics, in the process
of interaction with a detector, a quantum system exhibits particle properties,
hence, it could not be detected simultaneously by two different detectors. In
fact, the starting point was the analysis of the two slit experiment by N.
Bohr. This analysis played a fundamental role in elaboration of the complementarity
principle. In the two slit experiment by placing detectors directly
behind slits one does not observe coincidence detections, detectors never click
simultaneously.

Of course, this is the idealization of the real physical situation. The first
basic assumption is that the used source is really {\it a single-photon source.} At
least in 1920th such sources did not exist; even nowadays one can only approximately
assume that a source is of the one-photon type (with sufficiently
high approximation). 
Another problem is noise. In any event, there are coincidence
detections and their number is not negligibly small. Nevertheless,
it has to be relatively small.

The corresponding experiments have been done. The first experiment was
done by Grangier [41], [42], see also [43] for a review, in the framework of
quantum optics: detection of the outputs of two channels of the polarization
beam splitter (PBS). \footnote{Up to author's knowledge the experiment on the coincidence detection in the real two
slit experiment has never been done, at least with high precision.
} By quantum mechanics a single photon passing PBS
could not be split and coincidence detection for two detectors placed in the
two outputs of PBS is impossible. Grangier demonstrated that the relative
probability of coincidence detection, the {\it second order coherence}:
\begin{equation}
\label{E6}
g^{(2)}(0) =\frac{P_{12}}{P_1 P_2}
\end{equation}
is less than 1. At the same time all known (semi)classical field models predicted
that this coefficient exceeds 1, see [43] for review.

In fact, the von Neumann projection postulate (for observables with nondegenerate
spectrum) is the formal Hilbert space description of the absence
of coincidence detections. As the result of quantum measurement only one
detector can click. (We state again that we consider only noncompound
systems.) This physical process is represented in the operational quantum
formalism by projection of the state $\phi$  onto one fixed state $\phi_{k_0}.$ The quantum
collapse of superposition differs from the ``classical collapse" of the waves
superposition by its nondegeneration, the output is only a single state $\phi_{k_0}.$
Hence, in quantum theory only a single result of measurement, say $\omega_{k_0},$
can be obtained. The state  $\psi$ of the quantum system is ``collapsed" to the
eigenstate  $\phi_{k_0}.$ (Opposite to classical wave theory, all states are assumed to
be normalized.) In the mathematical formalism this process is described by
the von Neumann projection postulate:
\begin{equation}
\label{E7}
\psi_{k_0} = \frac{P_{k_0} \psi}{\Vert P_{k_0} \psi\Vert}
\end{equation}
The probability to get the result $\omega_{k_0}$ is given by Born's rule
\begin{equation}
\label{E8}
P(\omega_{k_0})
=\vert\langle \psi\vert \psi_{k_0} \rangle  \vert^2. 
\end{equation}
We remark that in quantum theory this rule is a postulate, i.e., it is not
derived from other natural physical assumptions. And there is no satisfactory
derivation of this rule from other (``naturally justified principles"), in spite
of numerous attempts to do this.\footnote{``The conclusion seems to be that no generally accepted derivation of the Born rule
has been given to date, but this does not imply that such a derivation is impossible in
principle". See [44]; cf. `t Hooft [45], [46] and Hofer [10].}

In classical wave theory, in principle, it is possible to measure simultaneously
all values $\omega_k.$ (We stress the role of simultaneous measurement. We
shall come back to this crucial point later.) Measurements induce extractions
of the components $\Phi_k$ which are proportional to eigenwaves $\phi_k$ from
the coherent superposition $\phi.$ see (3):
\begin{equation}
\label{E9}
\Phi_k = P_k \phi.
\end{equation}
Opposite to the quantum case this components are not normalized; the quantity
\begin{equation}
\label{E10}
{\cal E}_k =\Vert \Phi_k \Vert^2= \vert\langle \phi \vert \Phi_k  \rangle  \vert^2
\end{equation}
is the energy of the $k$th output wave, corresponding to the result of measurement
$\omega= \omega_k.$

The common between the quantum and classical cases is that in any
event the {\it coherent superposition is destroyed.} In the quantum case, it is
transformed into one of eigenstates, in the classical case into decoherent
mixture of eigenwaves $\phi_k.$ We remark that in the quantum case if the result
of the fixed measurement, $\omega = \omega_k,$  is not selected, then we also get decoherent
mixture of eigenstates:
\begin{equation}
\label{E11}
\rho=\sum_k P(\omega_k) P_k.
\end{equation}
We point out that in the classical framework the Born rule for quantum
probabilities can be formally written by using relative energies of the
classical output waves:
\begin{equation}
\label{E12}
 \frac{{\cal E}_k}{{\cal E}_\phi}
\end{equation}
where ${\cal E}_\phi$ is the total energy of the field. We state again that in the classical
wave $\phi$ is not normalized. If now we use the normalized (by its total energy)
wave, i.e.,
\begin{equation}
\phi \to \psi =\frac{\phi}{\Vert \phi \Vert},
\end{equation}
we see that the formal mathematical expression for classical relative energy
of the $k$th output channel coincides with (8), the Born's rule.\footnote{We suspect that our consideration is similar to Born's motivation of the rule for
probabilities (8); the motivation which he did not present in his paper [47]. This is a pity
that Born did not present any motivations for the rule (8). This formal postulation of the
basic probabilistic rule of quantum theory makes the impression of something completely
new and even mysterious.} Consider the
classical wave normalized by its total energy, i.e., the wave $\psi.$ In this case
the output wave is the mixture of normalized waves. It can be written in the
same way as (11) by using relative energies, instead of probabilities:
\begin{equation}
\label{E13}
\rho=\sum_k  \frac{{\cal E}_k}{{\cal E}_\phi} P_k.
\end{equation}

Heuristically it is clear that the number of detector's clicks in $k$th channel
is proportional to the energy distributed to this channel. However, it is not
a trivial task to find a class of classical waves which  produce quantum
statistics of clicks given by the rule:
\begin{equation}
\label{E14}
 P(\omega_k)= \frac{{\cal E}_k}{{\cal E}_\phi}.
\end{equation}
It is evident that waves have to be random and detectors have to be
(similar to quantum detectors) of the threshold type -- to produce individual
clicks and not continuous output signals. Another crucial constraint is that
there should be no coincidence detections. Hence, the temporal structure
of the random field has to be selected in such a way that, although the
field is continuous and so to say it is present everywhere -- in any detector,
simultaneous clicks (corresponding to random energy spikes) would occur
relatively rarely.

This is a good place to compare our model, PCSFT, with the De Broglie's
{\it Theory of the Double Solution.} By the latter quantum mechanics can be considered
as an operational formalism for a deeper theory, theory of classical
physical waves containing singularities. The singularities in waves correspond
to quantum particles. In PCSFT, particles do not present at all. Singularities
appear only at the level of measurements, as clicks of the threshold
type detectors. However, the common point is that PCSFT-fields are also
singular; these are stochastic processes valued in the $L_2$-space. So, the prequantum
fields are nonsmooth and discontinuous for almost all values of a
random parameter. 

This is also a good place to make a remark about nonlocality.
The Theory of the Double Solution is closely related to Bohmian
mechanics. The latter is definitely nonlocal, since the quantum potential is
nonlocal. On the other hand, PCSFT is a local classical field theory, see
[29]--[36]. In PCSFT a sort of classical nonlocality can be assigned to the
background random field. This field is not considered in the present paper,
it started to play a crucial role in classical wave modeling of compound
systems which we do not study in this paper, see [33]--[35]. However, such a
background field is a classical field, as in stochastic electrodynamics [48], and
has nothing to do with nonlocality of the Bohmian type. There is a rather
common opinion that creation of Bohmian mechanics was the improvement
of the Theory of the Double Solution, so nonlocality was also firmly incorporated
in De Broglie's views. However, the real inter-relation between views
of De Broglie and Bohm is more complicated. In particular, by reading the
De Broglie's book [27] I was not able to find any trace of nonlocality. In the
description of a compound quantum system, De Broglie [27], p. 74-76, of
course, used the nonlocal quantum potential induced by the wave function
of a compound system. However, the wave function by itself and hence the
potential corresponding to it were treated as related to the formal mathematical
description. The propagations of real physical waves corresponding
to a compound system are also coupled, but through random fluctuations of
the field of random media, a kind of the background field. In this sense De
Broglie's views are close to the views of the author of this paper.
I also stress that both for De Broglie and for me the starting point of
model's creation was Einstein's attempt to create a purely field model of
physical reality, see, especially, Einstein and Infeld [49]; cf. De Broglie [27],
p. 43: ``My first attempts to interpret wave mechanics in terms of the Theory
of the Double Solution in 1926-1927, were undoubtedly suggested to me
by Einstein's work on general relativity. Einstein believed that the physical
world should be described wholly by means of fields, well defined at every
point of space-time and obeying well-defined equations of a non-random
nature." As we shall see in section 3, the later statement is not valid for
PCSFT, we shall operate with random fields. However, ``late De Broglie"
also rejected ``naive Einsteinian determinism" and stressed the role of the
random background field, see [27], p. viii.

\section{Threshold detection of classical random fields}

We consider a threshold type detector with the threshold ${\cal E}_d.$ It interacts with a
random field $\phi(s; \omega),$ where $s$ is time and $\omega$ is a chance parameter describing randomness. For a moment, we consider the ${\bf C}$-valued random field (complex stochastic process). Later we shall consider random fields valued in finite and infinite dimensional complex Hilbert spaces. The finite dimensional case corresponds to detection
of internal degrees of freedom such as, e.g., polarization. We stress that the real physical situation corresponds to random fields with infinite-dimensional state space,
e.g., $H = L_2({\bf R}^3);$ the space of complex valued fields $\phi: {\bf R}^3  \to {\bf C}$ (or ${\bf C}^k$).
The energy of the field is given by ${\cal E}(s; \omega) =  \vert \phi(s; \omega) \vert^2$  (hence, the random field
has the physical dimension $\sim
\sqrt{\rm{energy}})$. A threshold detector clicks at the first
moment of time $\tau= \tau(\omega),$  when field's energy ${\cal E}$ exeeds the threshold:
\begin{equation}
\label{E3}
{\cal E} (\tau(\omega), \omega) \geq {\cal E}_d.
\end{equation} 
In the mathematical model the detection moment is defined as the {\it first hitting
time} [50]
\begin{equation}
\label{E4}
\tau(\omega) = \inf\{s \geq 0: {\cal E} (\tau(\omega), \omega) \geq {\cal E}_d\}.
\end{equation}

We consider the following detection scheme. After arriving to a threshold type
detector a classical random field behaves inside this detector as the Brownian motion in the space of (complex) fields. Thus $\phi(s, \omega)$ is the Wiener process: the Gaussian process having zero average at any moment of time $E\phi(s, \omega)=0$; and the
covariance function
\begin{equation}
\label{E5}
E\phi(s_1, \omega) \overline{\phi(s_2, \omega)}
 = \min(s_1, s_2) \sigma^2; 
\end{equation} 
in particular, we can find average of its energy
\begin{equation}
\label{E6}
E{\cal E} (s, \omega)=\sigma^2 s.
\end{equation}
From this equation,
we see that the coefficient
$\sigma^2 = \frac{E{\cal E} (s, \omega)}{S}$ has the physical dimension of {\it power.}
We are interested in average of the moments of the ${\cal E}_d$-threshold detection for
the energy of the Brownian motion. Since moments of detection are defined formally
as hitting times, we can apply theory of hitting times for the Wiener process, see
[50]:
\begin{equation}
\label{E7}
\bar{\tau}\equiv E \tau = \frac{{\cal E}_d}{\sigma^2} 
\end{equation}
or
\begin{equation}
\label{E8}
\frac{1}{\bar{\tau}}=\frac{\sigma^2}{{\cal E}_d} \;.
\end{equation}
Hence, during a long period of time $T$ such a detector clicks $N_\sigma$-times, where
\begin{equation}
\label{E9}
N_\sigma\approx \frac{T}{\bar{\tau}}=\frac{\sigma^2 T}{{\cal E}_d} \;.
\end{equation}
Consider now a random $\phi(s, \omega)$ valued in the $m$-dimensional complex Hilbert
space $H,$ where $m$ can be equal to infinity. Let $(e_j)$ be an orthonormal basis in $H.$
The vector-valued $\phi(s, \omega)$ can be expanded with respect to this basis
\begin{equation}
\label{E10}
\phi(s, \omega) = \sum_j \phi_j(s, \omega) e_j, 
\end{equation}
where $\phi_j(s, \omega)= \langle \phi(s, \omega)\vert e_j\rangle.$
This mathematical operation is physically realized as splitting of the field $\phi(s, \omega)$ 
into components $\phi_j(s, \omega)$ These components can be processed through mutually
disjoint channels, $j = 1, 2,...,m.$\footnote{We consider separation of a signal into disjoint channels as a part of the measurement
procedure. These channels correspond to separate detectors that are looking at the input state through
some kind of beam splitter. And it is assume that an input state is a single mode photon state.
(Hence, aforementioned channels do not correspond to different photon modes.)
} We now assume that there is a threshold detector in
each channel, $D_1, ..., D_m$ We also assume that all detectors have the same threshold
${\cal E}_d > 0.$

Suppose now that   $\phi(s, \omega) $  is 
the Wiener process valued in $H$. 
This process is determined by the covariance
operator $B: H \to H.$ Any covariance operator 
is Hermitian, positive, and trace-class  and vice versa. The complex Wiener process is characterized
by Hermitian covariance operator. We have, for $y\in H,$
$
E \langle y, \phi(s, \omega )\rangle =0, 
$ 
and, for $y_j\in H, j=1,2,$ 
$E \langle y_1, \phi(s_1, \omega )\rangle \langle \phi(s_2, \omega ), y_2 \rangle =\min(s_1,s_2) \langle B y_1, y_2 \rangle.
$ 
The latter is the covariance function of the stochastic process; in the operator form:
$
B(s_1, s_2) = \min(s_1,s_2) B.
$
We note that the dispersion of the $H$-valued Wiener process (at the instant of time $s)$ is given by 
$\Sigma^2(s)= E \Vert \phi(s, \omega) \Vert^2= s \rm{Tr} B.
$
The quantity ${\cal E}(s, \omega)= \Vert \phi(s, \omega) \Vert^2$ 
is the total energy of the Brownian motion signal at the instant of time $s.$ Hence, the quantity
\begin{equation}
\label{E11}
\Sigma^2\equiv \frac{\Sigma^2(s)}{s} = \rm{Tr} B
\end{equation} 
is the average power of this random signal. We stress that the average power is time-independent.  

We also remark that by normalization of 
the covariance function for the fixed $s$ by the dispersion 
we obtain the operator,
\begin{equation}
\label{E12}
\rho= B(s,s)/\Sigma^2(s)= B/ \rm{Tr} B,
\end{equation} 
which formally has all properties of 
the {\it density operator} used in 
quantum theory to represent quantum states.
Its matrix elements have the form $\rho_{ij} = b_{ij}/\Sigma^2.$ These
are dimensionless quantities. The relation (\ref{E12}) plays a fundamental role in our
approach : {\it each classical random process generates a quantum state (in general
mixed) which is given by the normalized covariance operator of the process. }One
can proceed the other way around as well: each density operator determines a class
of classical random processes.

Consider components $\phi_j(s, \omega)$ of the vector valued signal $\phi(s, \omega).$
Then $$ 
E \phi_i(s, \omega)\overline{\phi_j(s, \omega)}= \min(s_1,s_2) \langle B e_i, e_j \rangle= b_{ij}.$$
In particular, 
$\sigma_j^2(s) \equiv  E {\cal E}_j(s,\omega)\equiv E \vert  \phi_j(s, \omega)\vert^2= s b_{jj}.
$
This is the average energy of the $j$th component at the instant of time $s.$ We also consider 
the average powers of components
\begin{equation}
\label{E13}
\sigma_j^2 \equiv \frac{\sigma_j^2(s)}{s}= b_{jj}.
\end{equation}
We remark that the average power of the total signal is equal to the sum of  the average powers of its components.
\begin{equation}
\label{E14}
\Sigma^2= \sum_{j} \sigma_j^2.
\end{equation}
Consider now a run of experiment of the duration $T.$ The average number of clicks
for the $j$th detector can be approximately expressed as
\begin{equation}
\label{E15}
N_j \equiv N_{\sigma_j}\approx \frac{\sigma_j^2 T}{{\cal E}_d}.
\end{equation}
The total number of clicks is 
(again approximately) given by 
$N= \sum_j N_j \approx \frac{\Sigma_j^2 T}{{\cal E}_d}.$
Hence, for the detector $D_j,$  the probability of detection can be expressed as
\begin{equation}
\label{E16}
P_j \approx \frac{N_j}{N} \approx \frac{\sigma_j^2}{\Sigma_j^2} = \rho_{jj}.
\end{equation}
This is, in fact, {\it the Born's rule for the quantum state $\rho$  and the projection operator
$\hat{C}_j = \vert e_j\rangle \langle e_j\vert$  on the vector $e_j.$}  For the detector $D_j,$  the probability of detection can
be expressed as
\begin{equation}
\label{E17}
P_j = \rm{Tr} \rho \; \hat{C}_j. 
\end{equation}

\section{Coincidence counts}

The relative number of coincidence counts is given by second order coherence:
\begin{equation}
\label{E18}
g^{(2)}(0)= \frac{P_{12}}{P_1 P_2}
\end{equation}
Probabilities $P_j, j=1,2,$ for singles were found in (\ref{E17}); they coincide 
with corresponding quantum probabilities. For a single photon source, the $P_{12},$
the probability of clicks in both channels after a beam splitter, equals to zero. But, 
of course, nobody has yet seen really single photon sources. Therefore in reality  $P_{12}$
differs from zero. Our threshold detection model definitely contradicts to quantum mechanics
as a theory: the coincidence probability is positive (in any case for $P_1=P_2=1/2).$ 
However, we are not disappointed by this situation. We are sure that quantum mechanics is just 
an approximate model to describe probabilistic data. Moreover, this model cannot take into account 
experimental technicalities, such as e.g. the size of the discrimination threshold. Therefore its theoretical
prediction $g^{(2)}(0)=0,$ for ``single photon sources'', is far from the real experimental data. 
Nevertheless, even with the aid of this rough prediction it was possible to discard the semiclassical model.
The latter predicts that $g^{(2)}(0) \geq 1.$ And by getting $g^{(2)}(0)$ sufficiently small,  experiments claim
that their data match with the predictions of quantum mechanics and mismatch with predictions of the semicalssical 
model. By taking into account the above discussion, we have to test our model contra experiment and not contra the theoretical 
quantum formalism.   

We were not able to obtain a formula for $P_{12}$ in the framework of classical Brownian model fro the prequantum field.
We were able only to obtain an estimate from above, see  [37]. This estimate shows that in some range of variation of  
the detection threshold ${\cal E}_d$ the second order coherence $g^{(2)}(0)$ decreases with the increase of ${\cal E}_d.$ 
Hence, $g^{(2)}(0) < 1$ for sufficiently large ${\cal E}_d.$ Thus our model does not say that there are no coincidence counts,
neither that their number is relatively small irrelatively to magnitudes of experimental parameters. However, by our model these parameters can always be selected in such a way that detected data would be described (with a good approximation) by the quantum probabilistic formalism. Thus Devil is really in detectors and experimental technicalities.

Unfortunately, it seems that Grangier's type experiments with detailed monitoring
of dependence of the coincidence probability on the value of the threshold
have never been done.

\medskip

{\bf Conclusion.} By following L. De Broglie we stressed the formal analogy
between the operational descriptions of the processes of measurement of
classical waves and quantum systems, including the wave collapse. Our
analysis showed that the main experimental difference between measurements
of classical and quantum waves is the absence of coincidence detections in
the latter. Then we presented the classical wave model which reproduces the
quantum detection probabilities (described by the Born's rule) and at the same
time the number of coincidence detections is relatively small (for sufficiently
large detection threshold). 

\subsection{Acknowledgments}

I would like to thank Borje Nilsson and Sergei Polyakov for numerous discussions and constructive criticism.
This paper was written under support of the grant ``Mathematical Modeling of Complex Hierarchic Systems'' of the faculty of 
Natural Science and Engineering of Linnaeus University.

\end{document}